# Ultrafast Tracking of the Spallation Layer in Bulk Gold, Aluminum, and Steel

Nicolas Thomae[1, †], Julian Vollmann[1, †], Julian Freundel[1], Maximilian Spellauge[1], David Redka[1], Heinz P. Huber[1, *]

[1]Laser Center HM, Hochschule München University of Applied Sciences, Lothstr. 34, 80335 Munich, Germany

[†]Authors contributed equally

[*]Corresponding author

**Nicolas Thomae,** Laser Center HM, Hochschule München University of Applied Sciences, Lothstr. 34, 80335 Munich, Germany, +49 89 1265-3676, nicolas.thomae@hm.edu, Orcid-ID: 0009-0005-2123-8784

**Julian Vollmann,** Laser Center HM, Hochschule München University of Applied Sciences, Lothstr. 34, 80335 Munich, Germany, +49 89 1265-3676, vollmann.julian@hm.edu, Orcid-ID: 0009-0007-5929-8074

**Julian Freundel,** Laser Center HM, Hochschule München University of Applied Sciences, Lothstr. 34, 80335 Munich, Germany, +49 89 1265-3676, freundel.julian@hm.edu, Orcid-ID: 0009-0002-6190-9835

**Dr. Maximilian Spellauge,** Laser Center HM, Hochschule München University of Applied Sciences, Lothstr. 34, 80335 Munich, Germany, +49 89 1265-3676, maximilian.spellauge@hm.edu, Orcid-ID: 0000-0003-1741-8537

**Dr. David Redka,** Laser Center HM, Hochschule München University of Applied Sciences, Lothstr. 34, 80335 Munich, Germany, +49 89 1265-3676, dredka@hm.edu, Orcid-ID: 0000-0002-7306-2232

**Prof. Dr. Heinz Paul Huber,** Laser Center HM, Hochschule München University of Applied Sciences, Lothstr. 34, 80335 Munich, Germany, +49 89 1265-1686, heinz.huber@hm.edu, Orcid-ID: 0000-0003-2444-9833

Extreme manufacturing with ultrashort-pulse (USP) lasers at the physical limit of precision and efficiency requires understanding ablation dynamics on the picosecond-to-nanosecond timescale.[1,2] Pump-probe reflectometry (PPR) provides direct access to photomechanical spallation through Newton ring (NR) interference,[3,4] but this signature vanishes when the spallation layer becomes optically opaque or the ablated material strongly attenuates the probe.[5,6] Here, we combine PPR with phase-sensitive interferometric pump-probe (PPI) measurements[7,8] to track the spallation layer in bulk steel, aluminum, and gold. PPI resolves the propagating layer even when the reflected probe signal is suppressed by >95%. Joint PPR/PPI analysis with transfer-matrix modelling (TMM)[9] yields spallation layer thickness, vapor layer absorption, and the layer disintegration times. These quantities are key determinants of the energy coupling of subsequent pulses in GHz burst processing.

In PPR, a pump pulse initiates the ablation and a temporally delayed probe pulse measures the spatially resolved surface reflectance, from which the relative reflectance change $\Delta R/R_0$ is obtained.[3,4,10] Near the threshold of ablation, stress-confined isochoric heating initiates a tensile rarefaction wave, resulting in the ejection of a liquid spallation layer that is several tens of nanometers thick.[4,11,12] As the spallation layer propagates, periodic thin-film interference produces NR in $\Delta R/R_0$,[3] which serve as a direct fingerprint of the spallation layer dynamics. At higher fluence, the liquid is heated into a metastable state close to the critical point, triggering phase explosion and the ejection of vapor, clusters, and droplets (vapor layer).[13] In PPR, the probe pulse is incoherently scattered and absorbed in the ablated material, which creates a strong drop in $\Delta R/R_0$.[5,6] Both optical fingerprints were recently used to experimentally validate molecular dynamics (MD) simulations of FeNi.[5]

However, once the spallation layer becomes optically opaque or the vapor layer is strongly absorbing, the interference contrast is attenuated and the NR vanish.[5,6] Both conditions are promoted by a large energy-penetration depth in metals, such as gold.[5,13] The absence of NR in these metals may therefore reflect a loss of interferometric contrast rather than the absence of spallation. Consistent with this interpretation, PPR could not confirm spallation in bulk gold and revealed only a strong reflectance drop near threshold,[14] whereas NR were observed by PPR in gold thin films.[4]

To overcome this limit, we combine PPR and PPI measurements[7,8] on industrially relevant bulk metals with increasing effective energy deposition depth $d_{eff}$: stainless steel (AISI 304, $d_{eff} \approx 15$ nm),[15] aluminum ($d_{eff} \approx 30$ nm)[15] and gold ($d_{eff} \approx 100$ nm).[16] PPR and PPI experiments were conducted by 1030 nm pump pulses with a pulse duration of 1 ps, focused at an incidence angle of 37.8° with a beam waist of $30\ \mu\mathrm{m}$ and probed at normal incidence using frequency-doubled 515 nm pulses with a duration of 300 fs. The peak fluence was set to twice the ablation threshold $F_{\mathrm{thr}}$. More details are given in the materials and methods.

In steel, $\Delta R/R_0$ resolves pronounced NR corresponding with a spallation layer with velocity of $1125\ \mathrm{m}/s$,[6,17–19] which remain clearly visible up to approximately 2 ns (Fig. 1a). The corresponding phase change $\Delta\phi$ undergoes repeated cycles up to approximately 1 ns, reflecting the continuously increasing optical path length as the spallation layer moves away from the surface. This measured velocity is consistent with previous PPR measurements[6,17–19] and MD simulations of FeNi[5] when accounting for the approximately linear dependence of spallation layer velocity with peak fluence.

As shown in Fig. 1b, for gold, $\Delta R/R_0$ drops to below -0.95 within the initial 100 ps and remains near this value throughout the full 3.5 ns observation window. No NR are visible at any delay or radius. Despite the strong attenuation of the reflectance, the phase remains clearly measurable. The $\Delta\phi$ map shows periodic phase cycles across the irradiated area corresponding to an upward propagating spallation layer. These oscillations persist up to the maximum delay of 3.5 ns, yielding a spallation layer velocity of $330\ \mathrm{m/s}$ , in good agreement with reported velocities for gold thin-film ablation.[20]

Aluminum exhibits a behavior analogous to that of gold, with a rapid decline of $\Delta R/R_0$ below -0.95 within the initial 100 ps. Once more, no NR are discernible, but $\Delta\phi$ displays oscillations corresponding to a spallation layer with velocity of $907\ \mathrm{m}/s$. In contrast to gold, the oscillations persist only up to 500 ps.

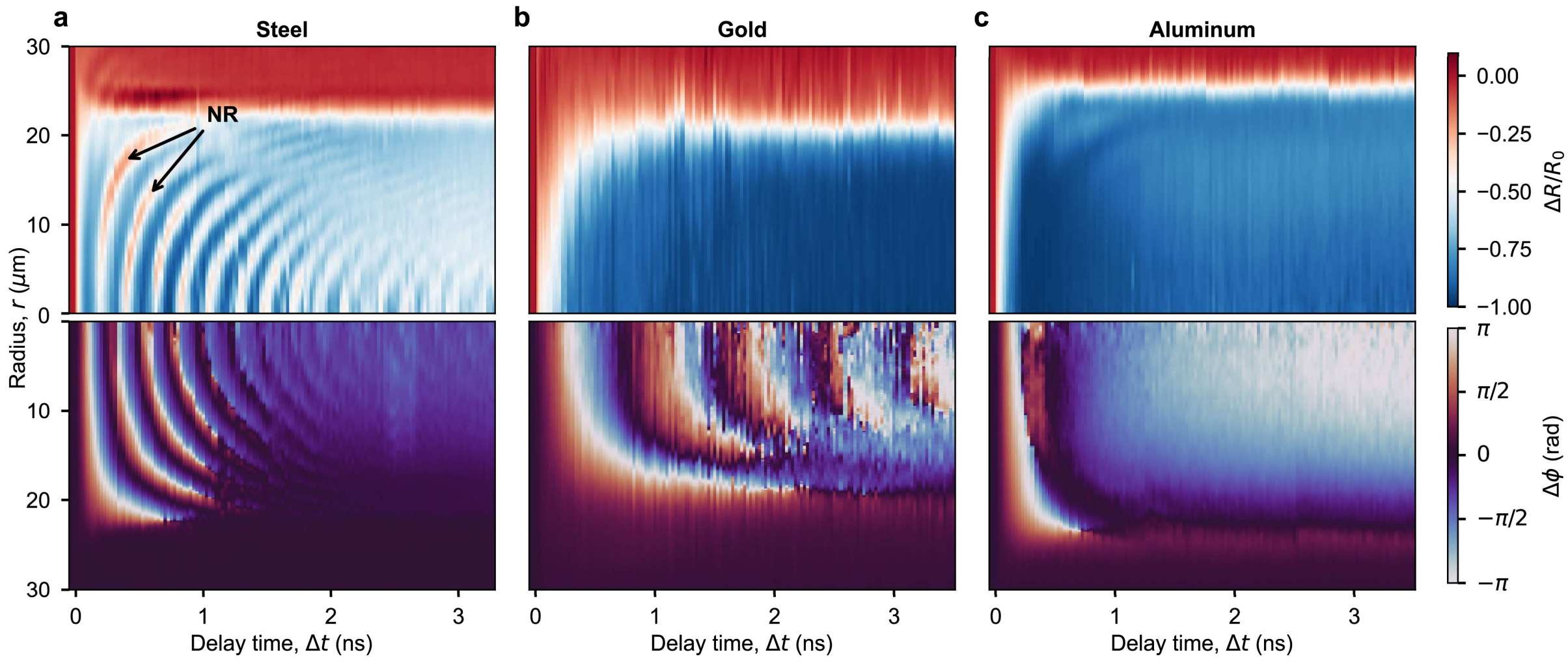

**Figure 1** Spatiotemporal maps of $\Delta R/R_0$ (upper row) and interferometric phase change $\Delta\phi$ (lower row) at an incident peak fluence of $2F_{\mathrm{thr}}$ for a) AISI 304 stainless steel, b) gold and c) aluminum. The maps show the pump-probe delay $\Delta t$ from 0 ns to 3.5 ns and the radial distance $r$ from 0 μm to 30 μm.

The PPI measurements resolve a moving spallation layer even when the reflected probe intensity is suppressed by more than 95% and NR are absent in PPR. To our knowledge, this is the first time-resolved tracking of the spallation layer in bulk, even though the probe beam is strongly attenuated. The absence of NR, however, can arise from two effects, an optically opaque spallation layer or strongly absorbing and scattering material.[6] To disentangle the two and resolve the transient material state above the surface, both $\Delta R/R_0$ and $\Delta\phi$ are analyzed simultaneously with TMM.[9] The multilayer system is motivated by the established theory of spallation as schematically shown at different delay times in Figure 2a. Details about the implementation and a sensitivity and identifiability analysis are given in the Supplement, Section 3.

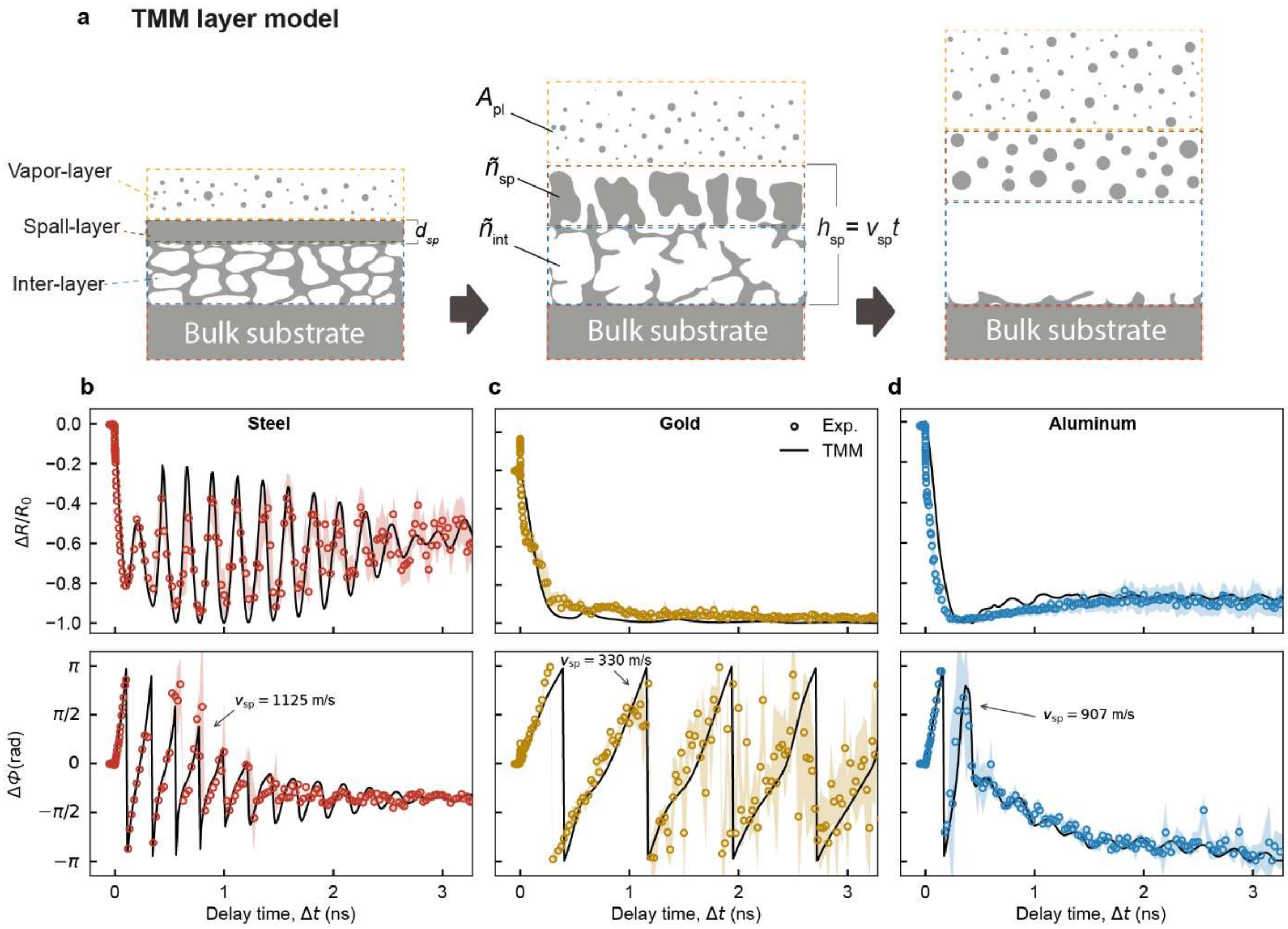

**Figure 2** TMM scheme and experimental PPR and PPI results for AISI 304 stainless steel, aluminum and gold. a) depicts the TMM model for three different delay times. The four-layer stack is applied to all three metals, comprising the bulk substrate with a complex refractive index $\tilde{n}$, inter-layer with a complex refractive index $\tilde{n}_{\mathrm{int}}(t)$, spallation layer with thickness $d_{\mathrm{sp}}(t)$ and complex refractive index $\tilde{n}_{\mathrm{sp}}$ and vapor layer with absorptance $A_{\mathrm{pl}}$. The velocity of the spallation layer $v_{\mathrm{sp}}$ is directly taken from the experiments. Below the scheme, the PPR extracted relative reflectance change $\Delta R/R_0$ (upper row, circles) and the PPI extracted phase change $\Delta\phi$ (lower row, circles), averaged over a $2\ \mu\mathrm{m}$ at the irradiation center, are shown for b) AISI 304 stainless steel, c) gold and d) aluminum, with the resulting traces of the TMM model shown as black lines.

The $\Delta R/R_0$ and $\Delta\phi$ transients for steel are shown in Figure 2b. The joint TMM analysis constrains the initial spallation layer thickness to 13 nm, below the optical penetration depth of 16 nm,[15] and the absorption of the vapor layer to 21%, both sufficiently low for NR to be visible in the $\Delta R/R_0$ signal. The complex refractive index of the liquid-bridge/void interlayer reduces the first NR maximum at 150 ps. After 200 ps the interlayer becomes vacuum-like ($\tilde{n}_{\mathrm{int}} \approx 1 - 0 \cdot i$) and the NR contrast is maximized. The fitted interlayer $\tilde{n}_{\mathrm{int}}$ closely reproduces $\tilde{n}_{\mathrm{int}}$ in FeNi derived from MD simulations (see Supplement, Section 2),[5] supporting the physical basis of the interlayer model. The NR persist about 2 ns in $\Delta R/R_0$. However, the contrast of the oscillation decreases progressively from the first NR period onward, which is explained by progressively increased amplitude of the field reflected from the bulk compared to the amplitude of the field reflected by the spallation layer. This is direct evidence of gradual optical thinning of the spallation layer as further analyzed in Ref. [[21]].

For gold, $\Delta R/R_0$ and $\Delta\phi$ transients as depicted in Figure 2c, are well fitted by an initial spallation layer thickness of at least 28 nm and velocity of 330 m/s with vapor layer absorption of 68 %. Above this thickness, the spallation layer is optically opaque, and the measured response becomes only weakly sensitive to further increases in $d_{\mathrm{spall},0}$. Thus, the TMM analysis provides a lower bound rather than a unique absolute thickness, as shown by the open residual minimum in the Supplement, Section 3. Even at this lower bound, the layer exceeds the optical penetration depth of 18 nm[22] and attenuates the lower-interface contribution required for NR to a round-trip transmission of only $\exp(-2 \cdot 28\,\mathrm{nm}/18\,\mathrm{nm}) \approx 0.04$ and is further suppressed by the 68 % absorption in the vapor layer resulting in 1.4 % of the lower-interface contribution, which prevents NR formation in PPR. In contrast, PPI remains sensitive to the front interface of the moving spallation layer. The phase oscillations persist with nearly constant amplitude beyond 3.5 ns, indicating that the spallation layer remains opaque throughout the observation window, with no measurable contribution from the underlying bulk surface.

For aluminum, $\Delta R/R_0$ and $\Delta\phi$ transients as depicted in Figure 2d, are explained with an initial spallation layer thickness of 20 nm, velocity of 907 m/s and vapor layer absorption of 66 %. As for steel, the localized joint residual minimum shows that the spallation layer thickness and vapor layer absorption are independently constrained by the combined observables. The spallation layer is only slightly thicker than the optical penetration depth of about 14 nm.[15] The bulk reflection is therefore attenuated, but not fully suppressed, with a round-trip transmission of about 5.7 %. The first destructive interference minimum, expected near 190 ps, indeed coincides with the minimum in $\Delta R/R_0$, but the following NR maximum expected near 340 ps is not observed. As shown in $\Delta\phi$, the oscillation of the spallation layer movement is only observed until 500 ps, indicating the disintegration of the spallation layer, which then no longer provides the reflection required for NR formation.

The fitted initial spallation layer thicknesses increase in the order steel (13 nm) < aluminum (20 nm) < gold (>28 nm). The layer thickness is expected to follow the effective energy penetration depth $d_{\mathrm{eff}}$, over which the absorbed energy is redistributed before electron-phonon equilibration. $d_{\mathrm{eff}}$ combines the optical penetration depth $d_{\mathrm{opt}}$ with the electron diffusion depth $d_{\mathrm{diff}}$ as $d_{\mathrm{eff}} = d_{\mathrm{opt}} + d_{\mathrm{diff}}$.[15,23] Within the two-temperature model,[11] $d_{\mathrm{diff}}$ is governed by the competition between the electronic thermal conductivity, which carries energy away from the surface, and the electron-phonon coupling strength, which dissipates it to the lattice.[24] A larger $d_{\mathrm{eff}}$ deepens the molten region and the zone of subsurface void nucleation and coalescence, so more molten material is ablated from the bulk, and a thicker spallation layer is ejected.[5,13,25] With $d_{\mathrm{eff}} = 15\ \mathrm{nm}$ for steel[15], $d_{\mathrm{eff}} = 30\ \mathrm{nm}$ for aluminum[15] and $d_{\mathrm{eff}} = 100\ \mathrm{nm}$ for gold[16], $d_{\mathrm{eff}}$ follows the order of the fitted initial spallation layer thicknesses.

In comparison to near-threshold MD predictions of approximately 10 nm for FeNi,[5] 50 nm for aluminum[25–27] and 100 nm for gold,[26,28] the steel value agrees, whereas the fitted gold and aluminum layers are markedly thinner. Our measurements were performed at $2\ F_{\mathrm{thr}}$, where a larger fraction of the overheated near-surface melt decomposes into vapor, clusters and droplets, thinning the spallation layer and feeding the vapor layer above it.[5,13,25] The fitted vapor layer absorption supports this independently, rising from 21 % in steel to 66 % in aluminum and 68 % in gold.

The decay of the phase-oscillation amplitude in the $\Delta\phi$ transients follows the spallation layer becoming progressively more transparent, providing time-resolved access to its optical thinning. The spallation layer disintegrates in aluminum after approximately 500 ps, in steel within 1 ns to 2 ns, whereas for gold it stays intact beyond 3.5 ns. Current MD simulations cannot yet show this lateral dissolution, which may be due to the limited lateral size (100 nm) and the periodic boundary conditions.[25,29,30] The measured lifetime establishes the timescale over which subsequent pulses in GHz burst processing interact either with an intact spallation layer or with disintegrated ablated material, thereby fundamentally altering energy coupling of a subsequent pulse.[31]

Our measurements thus reveal how the transient material state above the surface evolves and provide a direct experimental basis for understanding and optimizing GHz burst-mode processing. Resolving spallation layer lifetimes eliminates a fundamental limitation of established reflectometric methods, paving the way for greater precision and efficiency of USP processing, an extreme manufacturing technology operating at the physical limit.

## Materials and methods

### Samples and ablation thresholds

Measurements were performed on austenitic stainless steel (AISI 304), aluminum (99.999%) and gold (99.99%) with a RMS surface roughness $S_q$ of <3 nm, <4 nm and <5 nm for steel, aluminum and gold, respectively.

Ablation thresholds were determined with the $D^2$-method[32] from ten measurements per fluence, giving $F_{\mathrm{thr}}$ = 0.203(1), 0.46(1) and 2.0(1) $\mathrm{Jcm}^{-2}$ for steel, aluminum and gold, respectively (see Supplement, Section 1). These values agree with reported thresholds at comparable pulse durations for steel and aluminum[15] as well as for gold.[14,33]

### Pump–probe reflectometry and interferometry

A femtosecond regenerative amplifier (Pharos, Light Conversion, Lithuania) delivered 300 fs pulses at 1030 nm. The pump-pulse was stretched to 1.0 ps (Gaussian, FWHM), confirmed by autocorrelation (PulseCheck, APE GmbH, Germany). Single pump-pulses were selected by the internal pulse-picker and a triggered shutter and focused with a 100 mm plano-convex lens at an incidence angle of 37.8°, yielding a beam waist radius $w_0$ = 30(1) µm for steel and 29(1) µm for aluminum and gold (Gaussian, $1/e^2$), measured by a beam-caustic measurement device (MicroSpotMonitor, PRIMES GmbH, Pfungstadt, Germany). The frequency-doubled (515 nm) probe-pulse was delayed by a motorized translation stage and imaged at normal incidence in a Michelson-type interferometer through a long-working-distance objective (Plan Apo SL 50x, NA = 0.42, Mitutoyo, Japan), following the pump-probe microscopy design of Domke *et al.*[10] extended to imaging interferometry as in Temnov *et al.*[8] and Pflug et al.[22] Orthogonally polarized sample and reference beams were superposed by a 45° analyzer to generate spatial-carrier fringes with period of 1.9 um, which were recorded with a 14-bit CMOS camera (pco.pixelfly USB,

PCO AG, Germany). Time zero $\Delta t = 0$, defined as the point of maximum temporal overlap between pump and probe pulses, was calibrated on indium tin oxide via the transient reflectance peak originating from the Kerr effect as analyzed in Refs. [34,35]. For each delay, a pristine sample position was probed with two images: $R_0$ at $\Delta t \ll 0$, $R(\Delta t)$ at $\Delta t = -5$ ps to $\Delta t = 3.5$ ns and $R_\infty$ at $\Delta t = 1$s.

The relative reflectance change $\Delta R/R_0$, was evaluated pixel-wise as $\Delta R/R_0 = (R(\Delta t) - R_0)\ /\ R_0$ as done in Ref [10]. The resulting $\Delta R/R_0$ in the center was then averaged radially over 16 pixels corresponding to 2 μm, where the local fluence deviates by less than 1 % from the peak fluence.

The phase shift $\Delta\phi$ was retrieved from the interferogram by two-dimensional Fourier-transformation. The fringe period was set by the interferometer mirror to $\Lambda = 1.9$ μm. One AC sideband was isolated, shifted to the origin and inverse-transformed, with the argument giving the phase confined to $[-\pi, \pi]$. The relative phase shift was then given by $\Delta\phi = \phi(t) - \phi_0$. This method and evaluation were taken from Ref. [36]. This wrapped phase was analyzed directly, since spatial unwrapping is prone to artefacts from attenuated high spatial frequencies at sharp crater edges and from degraded fringe contrast by vapor layer absorption.[8]

Where an absolute displacement was required, $\Delta\phi$ was additionally unwrapped: phase maps were radially averaged using the complex mean over concentric elliptical contours,[8] and discontinuities between adjacent radii exceeding an empirically set threshold of $0.7 \cdot 2\pi$ were corrected by adding $2\pi$ (path-dependent unwrapping). Apparent displacement then follows $h = \Delta\phi\lambda/4\pi$; this relation is exact only for a purely geometric phase shift, as transient refractive-index changes also contribute to $\Delta\phi$. This contribution, however, only adds a few nanometers and is thus negligible.[18]

As a consistency check, $\Delta R/R$ was reconstructed from the PPI interferograms and compared with the independent PPR measurement showing excellent agreement (Supplement, Section 4).

**Transfer-matrix modelling**

Transient optical responses were calculated with a transfer-matrix model[9] comprising an absorbing vapor layer, a spallation layer with a time-dependent thickness, an effective liquid-bridge/void interlayer, and the bulk substrate. The vapor layer was parameterized solely by its absorption coefficient. The spallation layer thickness, vapor layer absorption and time-dependent interlayer complex refractive index were fitted to jointly reproduce the measured $\Delta R/R_0$ and the wrapped phase change $\Delta\phi$. The details about the implementation and used parameters are shown in the Supplement, Section 2.

Rather than statistical error bars, the fit is characterized by its sensitivity: varying the spallation layer thickness and vapor layer absorption around the optimum visibly degrades the agreement with both the reflectance and the phase, and the corresponding identifiability is documented in the Supplement, Section 3.

Complex refractive indices for the bulk substrate and spallation layer were taken from Ref. [18] for steel and aluminum and Ref. [37] for gold. The spallation layer velocity was directly taken from the experiments.

**Data availability**

Data are available on Zenodo and will be made public upon acceptance.

**Supporting Information**

Supporting Information accompanies the online version of this article and is available from the journal website.

**Acknowledgments**

This work was supported by Deutsche Forschungsgemeinschaft under Grant No. 528706678.

**Conflict of Interest**

The authors declare no competing financial or non-financial interests.

**Contributions**

**N.T.** and **J.V.** contributed equally to this work. **N.T.**: Writing – review & editing, Writing – original draft, Methodology, Investigation, Formal analysis, Validation. **J.V.:** Writing – review & editing, Writing – original draft, Methodology, Investigation, Formal analysis,

Validation, Visualization. **J.F.**: Methodology, Investigation, Formal analysis. **M.S.**: Writing – review & editing, Methodology, Validation. **D.R.**: Writing – original draft, Supervision, Investigation, Formal analysis. **H.P.H.**: Writing – review & editing, Writing – original draft, Project administration, Supervision, Conceptualization, Funding acquisition.